\documentclass[aps,prd, preprintnumbers, floatfix, superscriptaddress,showpacs]{revtex4}
\usepackage{graphicx}
\usepackage{amsmath,amsfonts,bm,bbold}
\usepackage{amssymb}
\begin{document}

\preprint{JLAB-THY-09-999}

\author{A.~V.~Radyushkin}
\affiliation{Physics Department, Old Dominion University, Norfolk,
             VA 23529, USA}
\affiliation{Thomas Jefferson National Accelerator Facility,
              Newport News, VA 23606, USA}
\affiliation{Bogoliubov Laboratory of Theoretical Physics, JINR, Dubna, Russian
             Federation}

\title{Shape of Pion Distribution Amplitude}

\begin{abstract}
 A   scenario is investigated  in which the leading-twist  pion distribution amplitude
$\varphi_\pi (x)$ is  approximated by the pion decay  constant $f_\pi$ for all essential values of 
the light-cone fraction $x$.
A  model for the light-front wave function $\Psi (x, k_\perp)$  is proposed that 
produces such a  distribution amplitude  and  has a rapidly decreasing  (exponential  for definiteness)  
dependence on the light-front  energy  combination
$ k_\perp^2/x(1-x)$.
It is   shown  that this model easily reproduces
 the fit of recent  large-$Q^2$ {\sc BaBar} data
on the  photon-pion transition form factor.
Some aspects of scenario with flat pion distribution amplitude are discussed.

\end{abstract}


\maketitle

\section{Introduction}

The pion distribution amplitude (DA) 
$\varphi_\pi (x)$ \cite{Radyushkin:1977gp,Lepage:1979zb}
is an important function  accumulating information about
momentum sharing between the quarks of the pion 
when the latter is in its valence $\bar q q$ configuration.
It is an inherent element of perturbative QCD calculations
of hard exclusive reactions involving the pion. 
From the solution \cite{Efremov:1978rn,Efremov:1979qk}, \cite{Lepage:1979zb} 
of the evolution equation for the pion DA it follows that independently 
of its 
 shape at low normalization point $\mu_0 \lesssim 1$\,GeV,
at large values of the probing momentum the pion DA 
acquires universal asymptotic 
form: \mbox{$\varphi_\pi (x, \mu \to \infty) \to 6 f_\pi x (1-x)$  \cite{Efremov:1978fi}.}
However, in practical calculations, it is  very important
to know what is the shape of the pion DA at moderate and low
scales $\mu$.  
The standard measure of the  width of the pion DA
is  the value $\langle \xi^2 \rangle$ of its second moment 
with respect to the relative momentum fraction 
variable $\xi = x - (1-x)$.  QCD sum rule \mbox{calculations 
\cite{Chernyak:1981zz}}  give large $\langle \xi^2 \rangle >0.4$
values for this moment (compared to $\langle \xi^2 \rangle =0.2$
for the asymptotic DA)  which indicates that the pion 
DA is  a wide function for $\mu^2 <1$\,GeV$^2$.
Recent lattice calculations \cite{DelDebbio:2005bg,Braun:2006dg} give
 $\langle \xi^2 \rangle \gtrsim 0.3$ 
for $\mu^2$ values in this region.
A direct calculation of the pion DA in the Nambu-Jona-Lasinio model
\cite{RuizArriola:2002bp} (see also   Ref.~\cite{Praszalowicz:2001wy}) 
 produces the result that $\varphi_\pi (x)=f_\pi$ for all values of 
the momentum fraction $x$,
i.e., that pion DA is constant.
The same  result  was obtained in the ``spectral'' quark model
\cite{RuizArriola:2003bs}.
The value of  $\langle \xi^2 \rangle $
for this ``flat'' DA  is 1/3, which is  compatible
with the results of the lattice estimates, though 
smaller than the result of QCD sum rules.
It  should be noted that the usual procedure of reconstructing 
pion DA from its moments in the CZ approach 
(followed by essentially all other groups)  is 
to build it as a sum of the lowest (two or three) Gegenbauer polynomials
corresponding to multiplicatively renormalizable 
components of the pion DA evolution decomposition.
Since these components have $x(1-x)$ as an overall factor,
such a procedure excludes flat DAs from possible models.
However, this restriction on the pion DA model building
is  just an assumption.
In the present paper, our goal is to analyze the photon-pion transition
form factor in a scenario with flat pion DA.
The curve for the form factor which we obtain
is  in a very good agreement with recent {\sc BaBar} 
 data  \cite{Aubert:2009mc},
which is  basically the main motivation for our investigation.

The paper is  organized in the following way. In Section II, 
we give an overview of the basic facts about the pion distribution
amplitude: its definition, evolution and results concerning  its shape.
Section III is  devoted  to the study of the photon-pion transition form factor.
We    briefly describe pQCD results for this form factor,
and then calculate it within the light-front formalism using
a model wave function $\Psi (x, k_\perp)$ that reproduces flat pion DA after 
integration over quark transverse momentum $k_\perp$
and rapidly (exponentially) decreases for large values 
of the standard light-front energy combination
$k_\perp^2/x(1-x)$.  The  $k_\perp$ width parameter $\sigma$ of 
this wave function can be easily adjusted to produce 
a curve practically coinciding with the data  fitting curve 
given in Ref.~\cite{Aubert:2009mc}.
This value of $\sigma$  corresponds to the value $\langle k_\perp^2 \rangle = (420\,{\rm MeV})^2$
for the average transverse momentum squared,
which has the magnitude that one would expect 
for the valence $\bar qq$ Fock component of the pion light-front wave function.
We analyze the structure of the one-loop corrections for a flat DA in pQCD,
and find out  that the optimal value $\mu^2 =aQ^2$ of the normalization 
scale for a flat DA is very small. We argue that this is an evidence 
that 
the flat pion DA  should  not be evolved in our calculation of the
photon-pion transition form factor. Finally, we discuss some aspects of the
flat pion DA scenario and then we summarize  the paper.

\section{Pion Distribution Amplitude: Basics}

\subsection{Definition  and Evolution}

The pion distribution amplitude $\varphi_\pi (x)$ may  be  introduced 
\cite{Radyushkin:1977gp}  as a function whose $x^n$ moments 
\begin{align}
f_n =  \int_0^1 x^n \, \varphi_\pi  (x) \,  dx 
\end{align}
are  given by reduced matrix elements of  twist-2 local  operators 
\begin{align}
i^{n+1}   \left \langle 0 | \bar d (0) \gamma_5 \left \{  \gamma_\nu D_{\nu_1} \ldots D_{\nu_n} \right \}  u (0) | \pi^+, P 
\right \rangle 
=  \left \{ P_\nu P_{\nu_1} \ldots P_{\nu_n} \right \} \, f_n  \  ,
\end{align}
or  \cite{Lepage:1979zb} as the $k_\perp$-integral   
\begin{align}
  \varphi_\pi  (x,\mu) = \frac{\sqrt{6}}{(2 \pi)^3} \, 
\int_{k_{\perp}^2\leq \mu^2}  \Psi(x,k_{\perp}) \, d^2 k_{\perp}
\label{phiLB}
\end{align}
of the light-front wave function  $\Psi(x,k_{\perp})$.
The zeroth moment of  $ \varphi_\pi  (x)$ corresponds to matrix 
element of the axial   current,  and  is  given by the pion decay constant $f_\pi$
\begin{align}
  \int_0^1  \varphi_\pi  (x) \,  dx = f_\pi \ , 
\label{fpi}
\end{align}
which is  known experimentally. In the conventions that we use,
$f_\pi \approx 130$\,MeV.  Eq.(\ref{fpi}) gives an important constraint
on the pion distribution amplitude (DA), fixing the integral under the
$ \varphi_\pi  (x)$ curve,  but it puts no restrictions 
on its shape. In fact, the pion DA depends on    
the renormalization scale $\mu$ that is  used to define 
matrix elements of  twist-2 local  operators: $  \varphi_\pi  (x) \to \varphi_\pi  (x, \mu)$.
The evolution equation for the pion DA may be written either in matrix form \cite{Radyushkin:1977gp}
\begin{align}
\mu \frac{d}{d \mu} \, f_n (\mu) = \sum_{k=0}^{n} Z_{nk} f_k (\mu) 
\end{align}
(see also \cite{Chernyak:1977as}) 
or in kernel form \cite{Lepage:1979zb}
\begin{align}
 \mu \frac{d}{d \mu} \varphi_\pi  (x, \mu) = \int_0^1 V(x,y)\, \varphi_\pi  (y, \mu) \, dy  \ .
\end{align}
The solution of the evolution equation was obtained \cite{Efremov:1978rn,Efremov:1979qk}, \cite{Lepage:1979zb} 
in the form of   expansion over Gegenbauer polynomials
\begin{align}
 \varphi_\pi  (x,\mu) = 6 f_\pi \, x (1-x) \,  \left \{ 1+ \sum_{n=1}^\infty a_{2n} C_{2n}^{3/2} (2x-1) 
\biggl [ \ln (\mu^2/\Lambda^2) \biggr ]^{-\gamma_{2n} / \beta_0} \right \} \  ,
\label{Gegen}
\end{align}
where $\gamma_{2n}>0$  is the anomalous  dimension  of the composite operator with $2n$ derivatives,
and $\beta_0$  is the lowest coefficient of the QCD $\beta$-function.
As a result,  when the normalization scale
$\mu$  tends to infinity, the pion DA acquires a simple form \cite{Efremov:1978fi}
\begin{align}
 \varphi_\pi  (x, \mu\to \infty) = 6 f_\pi \, x (1-x) \ ,
\label{asDA}
\end{align}
known as the ``asymptotic DA''. 

\subsection{Shape}

The question, however, is what is the shape of the pion DA
at low normalization scales $\mu \lesssim 1$ \,GeV. 
Some qualitative (and maybe overly simplistic  by today's standards) 
argumentation about   a possible shape of the pion DA was given 
in our 1980 papers \cite{Efremov:1980mb,Dittes:1981aw}.
  Namely, in  case of a system  of two  equal-mass non-interacting 
particles, 
  $ \varphi   (x) = \delta (x-1/2)$.
When the  interaction is switched on,  the DA broadens.
The width $\Gamma$ of $\varphi (x)$  may be estimated as 
$\Gamma \sim E_{\rm int} /m_q$.
Hence, for heavy mesons (e.g., for $\Upsilon$ particles),
 $ \varphi   (x) $ is  rather narrow since 
$m_q \gg M \sim  \Lambda_{\rm QCD}\sim 300$\,MeV,
where $M $ is  a  parameter characterizing 
interaction strength. 
On the other hand,   taking $m_{u,d} \lesssim 10$\,MeV
for the quarks in the pion, we  conclude that $\varphi_\pi  (x)$
is  very broad. Assuming a simple exponential model 
\begin{align}
 \Psi (x, k_\perp) \sim \exp \left [- \frac{k_\perp^2+m_q^2}{M^2 x (1-x)} \right ] 
\end{align}
for the light-front wave function gives 
\begin{align}
 \varphi_\pi  (x, \mu \sim M) \cong  f_\pi \exp \left [- \frac{m_q^2}{M^2 x (1-x)} \right ]  \ . 
\end{align}
In this case,  the pion DA $\varphi_\pi  (x)$
 is close to $f_\pi$  everywhere  outside the regions
$0\leq x \lesssim m_q^2 /M^2 \sim 10^{-3}$ and \mbox{$0\leq 1-x \lesssim m_q^2 /M^2 \sim 10^{-3}$.}
In these regions, $\varphi_\pi  (x)$ vanishes rapidly. 

\begin{figure}[ht]
\includegraphics[height=1.5in]{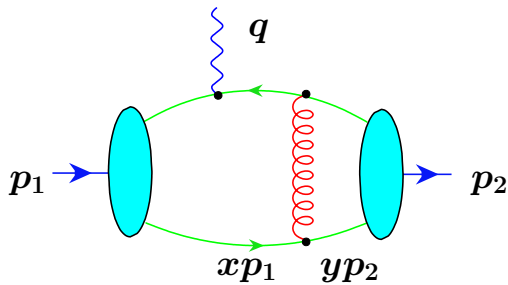}
\caption{
One-gluon-exchange diagram for pion electromagnetic  form factor 
in perturbative QCD. 
}
\label{piff}
\end{figure}

Initially,  the pion DA  appeared  in the  perturbative QCD 
expression \cite{Radyushkin:1977gp}
\begin{align}
 F_\pi^{\rm as \, (pQCD)}(Q^2)  = \frac{8 \pi \alpha_s  }{9Q^2} \int_0^1 dx \int_0^1 dy \
\frac{ \varphi_\pi  (x) \, \varphi_\pi  (y)}{xyQ^2} 
\end{align}
for the asymptotics of the pion form factor  calculated through a  one-gluon-exchange diagram (see Fig.\ref{piff}).
Here, $xyQ^2$ is  the virtuality of the exchanged gluon.
If one takes the flat pion DA $\varphi_\pi  (x) = f_\pi$, 
both integrals in $x$ and $y$ logarithmically diverge, which  means that
pQCD factorization fails in this  case. 
Evidently,  the finite size $R \sim 1/M$ of the pion should 
provide a  cut-off for the $x,y$ integral,  which suggests that 
$xyQ^2$ should be substituted by something  like
$xyQ^2 +{\cal O}(M^2)$, with the additional  term   having 
a meaning of the average of the squared transverse momentum of the quarks.
One may also treat ${\cal O}(M^2)$    as an effective gluon mass  squared.
Then the  $x,y$ integrals are convergent,  but the integral
is   dominated by the region  where the nominal gluon virtuality is 
 ${\cal O}(M^2)$. This means, first,  that it is small, and what is even more important, 
that it is 
not  growing with $Q^2$. 
For these reasons, the gluon-exchange line cannot be treated 
as a part of  a perturbative short-distance subprocess, 
in which virtualities of all the lines should 
be parametrically  ${\cal O}(Q^2)$.  Hence, it   should  be absorbed 
into the nonperturbative  part of the diagram, i.e. into  the soft 
pion wave function. The pion  form factor  must  be then calculated 
in some nonperturbative way.  Such  a  calculation was  accomplished 
within the QCD sum rule approach \cite{Ioffe:1982ia,Nesterenko:1982gc},
with the results  close to experimental data.

The   same QCD sum rule approach was used \cite{Chernyak:1981zz} 
to calculate
$\xi^2$  and $\xi^4$  moments of the pion distribution amplitude
$\phi_\pi (\xi)$ (which is the original DA $\varphi_\pi (x)$ written  as a function 
of the relative variable $\xi \equiv x -(1-x)$  and divided by $f_\pi$).
The  value of $\langle \xi^2 \rangle $ is a quantitative measure of the
width of the distribution amplitude.
In particular, $\langle \xi^2 \rangle $ is  zero for the infinitely narrow DA 
$\varphi_\pi (x) = f_\pi \delta (x-1/2)$,
it equals to 1/5 for the asymptotic DA (\ref{asDA})
and to 1/3  for the flat $\varphi_\pi (x) = f_\pi$ DA.
The  calculation of Chernyak and Zhitnitsky (CZ) \cite{Chernyak:1981zz} gave the result larger than 
$1/3$, namely 
$\langle \xi^2 \rangle=0.40 $
for the ``bare''  value that was attributed to the 
normalization point $\mu^2 = 1.5$\,GeV$^2$ and then 
renormalized to the  reference scale $\mu^2 = 0.25$\,GeV$^2$,
which resulted in  $\langle \xi^2 \rangle=0.46 $.
Without touching a subtle point whether a 
perturbative evolution to such a low scale is justified, we can say that 
the CZ  results  clearly 
indicate  that the pion DA is a wide function, and a generalized flat 
DA  of $\phi_\pi (\xi) =a + 3(1-a)\xi^2$
type could have been used as  a model  fitting the values of 
 $\langle \xi^2 \rangle$ and $\langle \xi^4 \rangle$
obtained from the CZ  calculation. 
However, the fitting model 
\begin{align}
 \phi_\pi^{\rm CZ} (\xi) =\frac{15}{4} \, \xi^2 \, (1- \xi^2)
\label{CZDA}
\end{align}
 was constructed from the sum of two first 
terms of the Gegenbauer expansion (\ref{Gegen}), 
which has $x(1-x)$ (or $1-\xi^2$) as an overall factor,
thus excluding all models with DA's that do not linearly vanish
at the  end-points. 

An implicit assumption of the CZ   calculation  is that it is  sufficient to take 
into account only the two 
lowest  condensates $\langle  GG \rangle$ and $\alpha_s \langle  \bar qq  \rangle^2$
in the operator product expansion (OPE) for the relevant two-point 
correlator.  
An  alternative attitude \cite{Mikhailov:1986be}  is that the quark  condensate 
$ \langle  \bar q(0) q(0)  \rangle$ is just the first term in Taylor expansion 
of the nonlocal condensate $ \langle  \bar q(0) q(z)  \rangle \equiv 
 \langle  \bar q q  \rangle \, f(z^2)$
that explicitly appears at the initial steps of  OPE calculations.
Modeling  $f(z^2)$  is  an attempt to 
include  the tower of  higher local condensates 
of  $\langle  \bar q (D^2)^n q  \rangle $ type.
The change from purely local approximation $f(z^2)=1$
to nonlocal condensates (NLC) with  a smooth function $f(z^2)$   that 
rapidly decreases  for large 
 $z^2$ modifies the QCD sum rule results for the moments 
of the pion DA: they become smaller.
 In particular, the initial NLC calculation  \cite{Mikhailov:1986be}
gave  $\langle \xi^2 \rangle=0.25$, and the model DA 
proposed in Refs.  \cite{Mikhailov:1986be,Mikhailov:1991pt}  is
\begin{align}
 \phi_\pi^{\rm MR} (\xi) = \frac{8}{\pi} \,  \sqrt{1-\xi^2} \  , 
\label{MRDA}
\end{align}
which is  wider 
than the asymptotic DA, but narrower than the flat DA.
The NLC  method  was elaborated in later papers, see Ref.  \cite{Bakulev:2007jv}
for a review.
The problem of the NLC approach is that while  it attempts to model 
the towers of $\langle  \bar q (D^2)^n q  \rangle $
condensates, the towers of $\langle  \bar q G^n q  \rangle $
condensates are neglected. Recently,   Chernyak \cite{Chernyak:2006ms}
gave a  specific example
in which the two towers exactly cancel each other. 
This means that  NLC results may underestimate 
the value of $\langle  \xi^2 \rangle$, and 
cannot exclude a possibility of   large 
$\langle  \xi^2 \rangle \gtrsim 1/3$ values for the second moment
and  $\langle  \xi^4 \rangle \gtrsim 1/5$ values for the fourth moment
of the pion DA. 
On the other hand,  it is   quite possible  that  NLC  argumentation  
is  not completely wrong, and   CZ  results overestimate the values of  
$\langle  \xi^2 \rangle$ and $\langle  \xi^4 \rangle$. In particular, 
recent lattice calculations \cite{DelDebbio:2005bg,Braun:2006dg}
give $\langle  \xi^2 \rangle \approx 0.29$ and 0.27, respectively,
at the scale $\mu^2=4$\,GeV$^2$,  which  produces 
$\langle  \xi^2 \rangle \gtrsim 0.3$  at scales $\sim 1$\,GeV$^2$,
but definitely not $\langle  \xi^2 \rangle \gtrsim 0.4$. 

A general comment  is that converting the obtained  values of  
$\langle  \xi^2 \rangle $ and $\langle  \xi^4 \rangle $
into models for the pion DA one should not restrict the models
by the requirement   
that  DA's  must  be given by  a few first terms 
of the  Gegenbauer expansion.  
There is no {\it a priori} principle justifying  such a  requirement:  it 
is just an assumption which  may or may not be true.

\section{Photon-pion  transition form factor}

The form factor $F_{\gamma^* \gamma^*  \pi^0}(q_1^2,q_2^2)$  relating two  
(in general, virtual)   photons with the lightest hadron, the pion, 
plays a special role in  the studies 
of  exclusive processes in quantum chromodynamics.
 When both photons are real, the form factor $F_{\gamma^* \gamma^*  \pi^0}(0,0)$
determines the rate of the $\pi^0 \to \gamma \gamma$ decay, 
and its value at this point is deeply related to the
axial anomaly \cite{anomaly}. 
At large photon virtualities, this  form factor  has 
the simplest structure analogous
to that of the  form factors
of deep inelastic scattering.
As  a result, comparing  pQCD predictions \cite{Lepage:1980fj,Lepage:1982gd,Brodsky:1981rp,delAguila:1981nk,Braaten:1982yp,Kadantseva:1985kb} 
with experimental data,
one can get  information about the shape 
of the pion distribution amplitude $\varphi_{\pi}(x)$.
Experimentally, 
 $F_{\gamma^* \gamma^*  \pi^0}(q_1^2, q_2^2)$ 
for small virtuality of one of the photons,
$q_2^2\approx 0$,  was measured at 
$e^+e^-$ colliders by CELLO \cite{Behrend:1990sr}, 
CLEO \cite{Gronberg:1997fj} and recently by {\sc BaBar} \cite{Aubert:2009mc}  collaborations.

\subsection{Perturbative QCD }

The 
behavior of photon-pion transition form factor at  large photon virtualities 
 was studied \cite{Lepage:1980fj,Brodsky:1981rp,Lepage:1982gd} within perturbative QCD (pQCD)
factorization approach for exclusive processes  \cite{Radyushkin:1977gp,Efremov:1979qk,Chernyak:1983ej,Lepage:1980fj}. 
Since  only
one hadron is involved, the
 $\gamma^* \gamma^*  \pi^0$ form factor has
the simplest  structure for pQCD analysis,  with
the nonperturbative information about the pion
accumulated in the pion distribution
amplitude $\varphi_\pi (x)$.  
Another  simplification is that  the short-distance
amplitude for $\gamma^* \gamma^* \to  \pi^0$  transition is given, 
at the leading order, just  by a single quark propagator.
Theoretically, most clean situation is when both photon virtualities
are large,  but  the  experimental study of
$F_{\gamma^* \gamma^*  \pi^0}(Q_1^2,Q_2^2)$  in this regime
through the  $\gamma^* \gamma^* \to \pi^0$ process
is very difficult due to very
small cross section.

\begin{figure}[ht]
\includegraphics[height=1.5in]{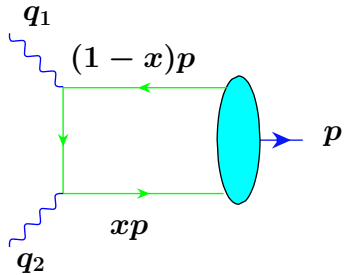}
\caption{Handbag diagram for photon-pion transition form factor.
}
\label{pigg}
\end{figure}

In the lowest order of perturbative QCD, the form  factor for  transition of {\it two}  virtual  photons
with momenta  $q_1,q_2$ into a 
neutral pion with  momentum $p=q_1+q_2$  is  given by  contribution of  the handbag  diagram (see Fig.\ref{pigg}) 
\begin{equation}  
F_{\gamma^* \gamma^*\pi}^{\rm pQCD,LO} (q_1^2, q_2^2) = -\frac{\sqrt{2}}{3} \int_0^1 
 \frac{\varphi_{\pi}(x)}{xq_1^2+(1-x)q_2^2} \, dx \  .
 \label{8a} 
\end{equation}
Introducing the asymmetry parameter $\omega$  through $q_1^2 = -Q^2 (1+\omega)/2$ and
$q_2^2 = -Q^2 (1-\omega)/2$  gives
\begin{equation}  
F_{\gamma^* \gamma^*\pi}^{\rm pQCD} (Q^2, \omega) =
 \frac{2\sqrt{2}}{3Q^2} \int_0^1 
 \frac{\varphi_{\pi}(x)}{1+\omega (2x-1) } \, dx \  \equiv  \frac{\sqrt{2}f_\pi}{3\, Q^2}\, J(\omega) \ .
 \label{8} 
\end{equation}
Thus,  if  one would  know the function $J (\omega)$, one could (in principle) obtain the pion DA
$\varphi_{\pi}(x)$  by inverting the integral transform (\ref{8}). 
However, as already mentioned, this kinematics is very difficult for experimental study. 
If one of  the photons is  real, i.e. $\omega=1$, the leading-order pQCD 
prediction is 
\begin{equation}  
F_{\gamma^* \gamma \pi}^{\rm pQCD} (Q^2) = 
\frac{\sqrt{2}}{3Q^2} \int_0^1 
 \frac{\varphi_{\pi}(x)}{x} \, dx \  \equiv  \frac{\sqrt{2}f_\pi}{3\, Q^2}\, J \ .
 \label{8b} 
\end{equation}
Information about the shape of the pion wave function is  now accumulated in the factor $J$.
It equals 2 for the infinitely narrow $\sim \delta (x-1/2)$ DA, for asymptotic DA (\ref{asDA})
we have $J^{\rm as}=3$, while CZ model (\ref{CZDA}) gives $J^{\rm CZ}=5$.
The  intermediate distribution (\ref{MRDA}) produces $J^{\rm MR}=4$. 
Thus, in addition to $\langle \xi^2\rangle $, we have another 
measure of  the width of the pion DA,  the value of $J$.
Note,  that for the DA's  listed above, the  ordering in $J$ values is the same 
as the ordering in $\langle \xi^2\rangle $ values.  
However, the flat DA, for which  $\langle \xi^2\rangle $ is  smaller than that 
for the CZ model DA, generates infinite value for $J$,
which is a consequence of the  fact that it does  not vanish at $x=0$.
This divergence of the integral for $J$ formally means  that the standard 
perturbative QCD factorization approach is  not applicable 
for the flat DA case.  
But, since the divergence is  only logarithmic, one  may
hope that some minimal  fix, like a cut-off, might be sufficient. 
The question, of course,  is whether there is a  real   need to use the flat DA 
to describe the data on the photon-pion  transition form factor.

\subsection{Logarithmic model}

\begin{figure}[ht]
\includegraphics[width=3in]{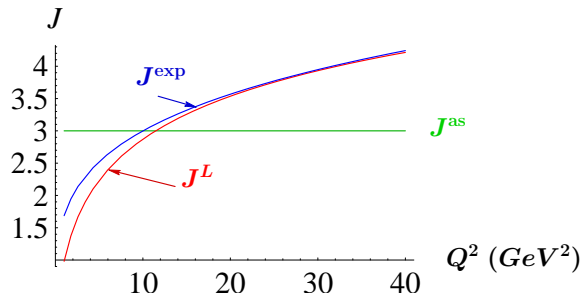}
\caption{Comparison of the function $J^{\rm exp} (Q^2)$ corresponding to 
 the fit of   {\sc BaBar} data (blue online)
and logarithmic   model function $J^{L} (Q^2) $ (red online).
  The asymptotic pQCD prediction $J^{\rm as}=3$ is also shown (green online). 
}
\label{fig1}
\end{figure}

Recent   data on $\gamma^* \gamma \to  \pi^0$ form factor reported by 
{\sc BaBar} collaboration in Ref.~\cite{Aubert:2009mc} are well  fitted by the formula
\begin{equation}
 Q^2 \,F_{\gamma^*\gamma \pi^0} (Q^2)  \cong  \sqrt{2} f_\pi \left  ( \frac{Q^2}{10\,{\rm GeV}^2}\right )^{0.25} 
\equiv \frac{\sqrt{2} f_\pi}{3} \, J^{\rm exp} (Q^2) 
\end{equation}
for the range 4\,GeV$^2<Q^2<40$\,GeV$^2$. 
The most startling observation is that $ J^{\rm exp} (Q^2)$  does not show a tendency 
to  flatten to some particular value.
The specific $(Q^2)^{\beta}$ power-law parametrization of the growth is, of  course,
a matter of  choice. 
In this region, $J^{\rm exp} (Q^2)$ is in fact  very close 
to  the logarithmic function
\begin{align}
 J^{L} (Q^2) = \ln \left (1+ \frac{Q^2}{M^2} \right )  \  ,
\end{align}
 if one takes $M^2=0.6$\,GeV$^2$,  
see Fig.~\ref{fig1}.
The two curves practically  coincide for $Q^2 \gtrsim 15$\,GeV$^2$. 

It is  easy to notice  that $ J^{L} (Q^2) $   can be   obtained 
if one uses the flat DA $\varphi_\pi (x) =f_\pi$ and 
changes $xQ^2 \to xQ^2+M^2$ in the pQCD expression for 
the $\gamma^*\gamma \to  \pi^0$ form factor:
\begin{align}
 J^{L} (Q^2) =  Q^2  \int_0^1 \frac{dx}{xQ^2+M^2} \ .
\label{JL} 
\end{align}
As discussed above, the idea to modify   propagators $1/k^2 \to 1/(k^2+M^2)$
 in integrals over the light-cone momentum fractions
is  rather old. The parameter $M$ in such modifications 
is usually treated as the  average transverse momentum
of the propagating  particle.
 However, the immediate  observation is that
 the value $M = 0.77$\,GeV is a little bit too large 
to be interpreted in such a  way.
Furthermore, the $1/xQ^2 \to 1/(xQ^2 + M^2)$ modification 
is equivalent to bringing in, {\it before the integration over $x$,}  a tower of $(M^2/xQ^2)^n$ power corrections,
i.e., higher twists.
But it is known \cite{Musatov:1997pu} that the handbag diagram, 
because of its simple singularity structure, cannot generate an  infinite 
tower of power corrections.
Indeed, the  propagator of a massless quark 
in the coordinate representation  is $\sim \not \! z (z^2)^{-2}$.
Expanding the matrix element of the bilocal operator
\begin{equation}
 \langle  0 | \bar \psi (0) \gamma_5 \not \! z  \psi (z) | p \rangle
=  \xi_2(zp)|_{z^2=0}  + z^2 \xi_4(zp)|_{z^2=0} + (z^2)^2 \xi_6 (zp)|_{z^2=0} + \ldots  \ ,
  \label{10}
 \end{equation}
we see that twist-6 and higher terms cancel the singularity  of the 
propagator. Hence, there are just two  terms in the OPE for the handbag contribution:
twist-2 term that has $1/Q^2$ behavior and twist-4 term 
(corresponding to the $\bar \psi  \gamma_5 \not \! z D^2 \psi $ operator on the light cone)
that gives $1/Q^4$ contribution. Operators with $(D^2)^{n\geq 2}$
do not contribute, and so there is  no infinite tower of 
$(1/Q^2)^n$ terms.

\subsection{Light-front  formalism and Gaussian  model}

\begin{figure}[ht]
\includegraphics[width=3in]{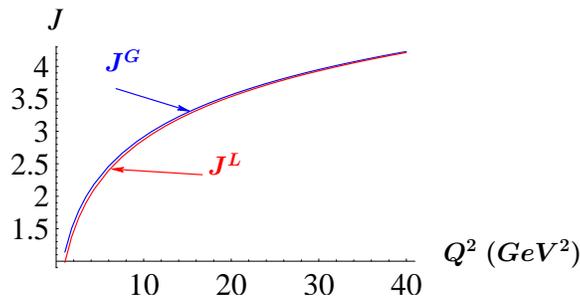} 
\caption{\label{psiphi}
Comparison of the logarithmic   model 
$J^L (Q^2, M^2=0.6\,{\rm GeV}^2)$  (red online)
and Gaussian  model  $J^G (Q^2,\sigma=0.53\,{\rm GeV}^2)$ (blue  online).
}
\label{fig2}
\end{figure}

\begin{figure}[ht]
\includegraphics[width=3in]{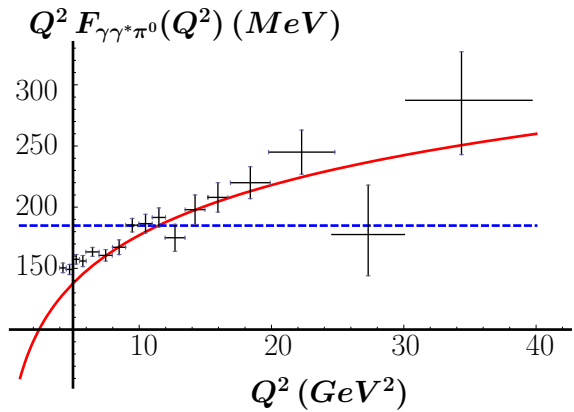} 
\caption{Comparison of model curve (solid, red online) with {\sc BaBar} experimental data.
  The asymptotic pQCD prediction 
$Q^2 F_{\gamma \gamma^* \pi^0}(Q^2)= \sqrt{2} f_\pi$ 
is also shown (dashed, blue online). 
}
\label{exp}
\end{figure}

To investigate  a  possible  mechanism capable of  generating a   cut-off  at small $x$,
let us write  the $\gamma^*\gamma \pi^0$
form factor in the light-front formalism.
The required expression was given in the classic paper  \cite{Lepage:1980fj}
of Lepage and Brodsky   on exclusive processes in QCD.
Namely,  the two-body ($i.e., \, \bar qq $) contribution 
to the $\gamma^*\gamma \pi^0$  form factor 
is given  by 
\begin{equation}
(\epsilon_{\perp} \times q_{\perp}) F^{\bar qq}_{\gamma^*\gamma \pi^0} 
(Q^2) = \frac{1}{4\pi^3 \sqrt{3}} 
\int_0^1 dx \int  
\frac{(\epsilon_{\perp} \times (xq_{\perp}+k_{\perp})) }
{ (xq_{\perp}+k_{\perp})^2- i \epsilon} \, \Psi(x,k_{\perp}) \,  d^2 k_{\perp} \, .
  \label{75} \end{equation}
Here, $q_{\perp}$ is a two-dimensional vector in the transverse plane
satisfying $q_{\perp}^2=Q^2$,   $\epsilon_{\perp}$ is a vector orthogonal to 
 $q_{\perp}$ and also lying in 
the transverse plane \cite{Lepage:1980fj}, and the cross denotes the vector product.
It   can be shown that  for   the wave functions of $\Psi(x,k_{\perp})= \psi(x,k_{\perp}^2)$
type we have \cite{Musatov:1997pu}
\begin{equation}
F^{\bar qq}_{\gamma^*\gamma \pi^0} 
(Q^2) = \frac{1}{2\pi^2 \sqrt{3}} 
\int_0^1 \frac{dx}{xQ^2} \int_0^{xQ}  
\, \psi(x,k_{\perp}^2) \, 
k_{\perp} d k_{\perp}
\, .
  \label{75A} \end{equation}
Following  \cite{Lepage:1982gd}, we  take the  Gaussian ansatz  for 
the  $k_\perp$-dependence of the  light-front wave function,
which we write in the form 
\begin{equation}
\Psi^G   (x,k_{\perp})  = \frac{4 \pi^2 \varphi_\pi (x)}{{x \bar x}\sigma \sqrt{6} } 
\,
 \exp \left (- \frac{k_{\perp}^2 }{2\sigma x \bar x} \right )
 \,  , 
\label{70} 
\end{equation}
where  $\sigma$ is the
width parameter  and $\varphi_{\pi}(x)$ is 
the desired  pion distribution amplitude.  
The result  for the form  factor  is  then  given  by 
\begin{equation}
F^{G}_{\gamma^*\gamma \pi^0}(Q^2) = 
\frac{\sqrt{2}}{3} \int_0^1 \frac{\varphi_{\pi}(x)}{x Q^2} \left [ 1- \exp \left 
( -\frac{xQ^2}{ 2\bar x \sigma } \right ) \right ] dx  \  .
\label{77}  
\end{equation} 
It    contains  the $1/xQ^2$ pQCD contribution   and a  correction term 
which makes the integral   convergent in the region of small $x$.
An important observation is that the correction term in the integrand of Eq.(\ref{77}) 
reflects the $k_\perp$  dependence of the 
nonperturbative pion wave function. In 
the Gaussian  ansatz,  this {\it integrand} term  
has an exponentially decreasing
rather than a power behavior for large $Q^2$.
This fact alone is  sufficient to assert that 
 it  cannot be classified as a higher-twist term.
It comes from  contributions invisible in the operator product expansion,
which only sees the terms that have a powerlike behavior in $1/Q^2$
{\it before integration over $x$}. 
Representing this  expression for  the   form factor as 
\begin{align}
 F^{G}_{\gamma^*\gamma \pi^0}(Q^2) 
= \frac{\sqrt{2}f_\pi}{3} \,  J^G(Q^2,\sigma) \ ,
\end{align}
we  find that,  for the flat DA $\varphi_\pi (x) =f_\pi$,  the function $J^G(Q^2,\sigma)$  has the following 
large-$Q^2$  asymptotic behavior: 
\begin{align}
  J^G(Q^2,\sigma) = \ln \left ( \frac{Q^2}{2 \sigma} \right ) 
+ \gamma_E + {\cal  O}(\sigma /Q^2) \  ,
\end{align}
where $\gamma_E$ is  the Euler-Mascheroni constant. 
Comparing this  result with the function $J^L (Q^2,M^2)$  (\ref{JL}) 
 obtained through the $M^2$ modification 
of the pQCD $1/xQ^2$ propagator,
we  conclude that they have the same (up to ${\cal O}(1/Q^2)$ terms) 
asymptotic behavior if 
\begin{align}
 \sigma = \frac{M^2}{2}\, e^{\gamma_E} \ ,
\end{align}
which gives $\sigma =0.53$\,GeV$^2$ for $M^2=0.6$\, GeV$^2$.
In fact, plotting $J^L (Q^2,M^2=0.6\,{\rm GeV}^2)$ 
and $J^G(Q^2,\sigma=0.53\,{\rm GeV}^2)$ together, we observe that 
these two functions practically coincide in the whole region 
$Q^2 >1$\,GeV$^2$ we are interested in (see Fig.~\ref{fig2}).
Comparison of the model curve with {\sc BaBar} experimental data is shown 
in Fig.~\ref{exp}.

To check  if the magnitude of $\sigma$ is in a physically reasonable range, 
let us  calculate the average transverse momentum 
for this  Gaussian model. We have
\begin{align}
\langle k_\perp^2 (x) \rangle  \equiv \int d^2 k_\perp  \, 
 k_\perp ^2 \Psi (x, k_\perp) \left (  \int d^2 k_\perp \, dx \, 
  \Psi (x, k_\perp) \right )^{-1} = 2 \sigma\, x(1-x) \  ,
\end{align}
and, hence, 
\begin{align}
 \langle k_\perp^2  \rangle  \equiv \int_0^1 \langle k_\perp^2 (x) \rangle \, dx 
 = \frac{\sigma}{3} \ .
\end{align}
Thus,  $\sqrt{\langle k_\perp^2  \rangle} = 0.42\, {\rm GeV}$,
which is   rather close to the folklore value of 300\,MeV.  
One should also take into account  that the wave function under  consideration 
describes  the valence two-quark 
Fock component of the pion, which is  presumably smaller than  other components.

Thus,  the magnitude of the $M^2$-parameter of the logarithmic model 
is  close to $3.3\, \langle k_\perp^2  \rangle $ rather than to the value $\langle k_\perp^2  \rangle $
 expected  from a naive substitution $xQ^2 \to xQ^2+ k_\perp^2$
in the  quark propagator. As we  explained, such a change has no 
theoretical  grounds 
in the case of the handbag diagram. The justification
  of the {\it ad hoc} modification  $xQ^2 \to xQ^2+ M^2$
used in our  logarithmic model, as we have seen, is   more complicated.

\subsection{One-loop pQCD corrections}

As already discussed, 
distribution amplitudes   in general depend on the  factorization 
 scale $\mu$, i.e. in principle 
one should always write:  $\varphi (x,\mu)$. This  dependence is  induced by radiative corrections.
The standard procedure in pQCD calculations involving pion DA is  to start with an auxiliary 
quark-antiquark state in which the quarks are on shell and share 
the total momentum $P$ in fractions $xP$ and $(1-x)P$ according to the  
 ``bare'' distribution amplitude $\varphi_0 (x,m_q)$.
Calculating radiative corrections for a specific process, e.g. for photon-pion
transition form  factor, one obtains logarithms $\ln (Q^2/m_q^2)$
accompanied by factors which  may be converted into convolution
of  the lowest-order short-distance amplitude 
$T_0 (x) $   with the  evolution kernel $V(x,y)$ and the bare
distribution amplitude $\varphi_0 (y,m_q)$. Combining the evolution factor with bare
DA, one obtains the  expression in which $T_0 (x) $  is multiplied by ``evolved''
distribution amplitude $\varphi (x, aQ)$, with  $a$ being  some number,  
which is  usually  chosen in such a way as  to minimize the size
of that part of the corrections which was not absorbed into the
renormalized (i.e. evolved) DA. 
One may also start with massless on-shell quarks, and use
dimensional regularization to regularize mass singularities
that result from  $\ln (Q^2/m_q^2)$ terms for $m_q =0$.
Then the bare DA  depends on  the dimensional regularization 
scale $\mu$, and one gets $\ln (Q^2/\mu^2)$ evolution 
logarithms calculating corrections to the amplitude of the 
 short-distance subprocess.

The one-loop correction  for the $\gamma^*\gamma \to \pi^0$
form factor was calculated in Refs.\cite{delAguila:1981nk,Braaten:1982yp,Kadantseva:1985kb},
with the result 
\begin{equation} 
\int_0^1 dx \, \frac{\varphi_\pi (x)}{ xQ^2}\,   \to 
\int_0^1 dx \, \frac{\varphi_\pi (x,\mu)}{ xQ^2} \biggl 
\{ 1+ C_F \frac{\alpha_s}{2 \pi} \biggl 
[ 
\frac1{2} \ln^2 x - \frac{x \ln x}{2(1-x)}  - 
\frac9{2} +\biggl (\frac3{2} + \ln x \biggr ) 
 \ln \left (  \frac{Q^2}{ \mu^2} \right )  \  \biggr ] \biggr \}
\equiv f_\pi \frac{J(Q,\mu)}{Q^2} \  . 
\label{14}
 \end{equation}
As advertised, 
 the term containing 
 the    logarithm $\ln (Q^2 / \mu^2)$  
has  the form of  convolution
\begin{equation}
\frac{1}{x Q^2} \, C_F \frac{\alpha_s}{2 \pi}
\biggl (\frac3{2} + \ln x \biggr )  =
\int \limits_0^1 \frac{1}{\xi Q^2} \, V(\xi, x)  \, d\xi 
\label{15} \end{equation}
of the lowest-order  term $T_0(\xi, Q^2) = 1/\xi Q^2$
and the  kernel 
\begin{equation}
V(\xi,x) = \frac{\alpha_s}{2 \pi}\, C_F \, 
\left [ \frac{ \xi }{ x} \, 
\theta(\xi <  x) \left ( 1 + \frac{1}{  x- \xi } \right )
+ \frac{1- \xi }{1-x} \,  \theta(\xi > x) 
\left ( 1 + \frac{1 }{\xi - x} \right )
\right ]_+ \label{16}
 \end{equation}
governing the evolution of the pion distribution amplitude.
The ``+''-operation is defined   by 
\begin{equation}
[F(\xi,x)]_+ = F(\xi,x) -\delta (\xi-x) \int \limits_0^1
 F(\zeta, x)\, d \zeta \,  . \label{17}
 \end{equation}

When the probing momentum $Q$ is   much larger
than the initial  normalization scale $\mu$,
one deals with large logarithm $\ln (Q^2/\mu^2)$.
The latter can  be eliminated by  taking $\mu=Q$,  and  
the expression is produced in which the {\it evolved}  DA $\varphi_\pi (x,Q)$ 
is integrated with the remaining part of the  correction.
It is  not  guaranteed, however, that the resulting correction will be small,
and the idea is to take $\mu = aQ$ with $a$ chosen in such a way as 
maximally reduce the size of the $\alpha_s$ correction.

In the context of the present paper, we are interested in 
what  happens when  the bare DA is flat:  $\varphi_0(x,\mu) = f_\pi$.
Since in this  case all integrals in (\ref{14}) simply diverge, let us take a  regularized 
version of the flat distribution amplitude, namely 
the function
\begin{align}
 \varphi_r (x) = f_\pi \, \frac{\Gamma (2+2 r)}{\Gamma^2 (1+r)} \, x^r (1-x)^r  \  , 
\end{align}
with $r$ being a very small parameter, say $r \lesssim 0.1$.
Then Eq.~(\ref{14}) 
gives
\begin{align}
 J_r(Q,\mu) = \left (\frac1{r} +2 \right ) \left \{ 1 + \frac{\alpha_s}{3\pi} \left [ \frac{2}{r^2} +\frac{\pi^2}{3} -9 +
{\cal O} (r) - \left ( \frac{2}{r} -3 + \frac{\pi^2}{3}\, r + {\cal O} (r^2) \right ) \ln \left ( \frac{Q^2}{\mu^2}\right ) 
\right ] \right \} \  .
\label{Jr}
\end{align}
It is  clear  that if  we take $\mu=Q$, we will  be left with 
a huge correction $\sim (2\alpha_s/3\pi)/r^2$,
i.e. $\sim 60 \, (\alpha_s/\pi)$ for $r=0.1$. Since the coefficient in front of  $\ln ({Q^2}/{\mu^2})$ 
is dominated by $2/r$ term, while the $\mu$-independent piece  is  dominated by its $2/r^2$ part,
we can compensate the latter by taking $\ln ({Q^2}/{\mu^2})=1/r$.
This corresponds to the choice
\begin{align}
 \mu^2 = Q^2 \, e^{-1/r} \ .
\label{1/r}
\end{align}
Thus, if we take $r=0.1$  to model the flat DA,
 the optimal choice for $\mu$ is something like $\mu^2 = 10^{-4}Q^2$.
Even for the highest $Q^2$ reached in {\sc BaBar}  experiment, this gives 
$\mu^2= 0.004$\,GeV$^2$, a scale corresponding to distances much larger
than the pion size.  Evidently, we cannot evolve the pion DA down to such small momentum scales.
The evolution must stop at some $\mu_0^2 \sim  \Lambda_{\rm QCD}^2$.
Thus, the flat pion DA  becomes
a DA at ``low normalization point''  $\mu = \mu_0 \sim \Lambda_{\rm QCD}$,  
below which there is no evolution.
Moreover,  as we have seen in the example above, the radiative corrections
do  not induce visible ${\cal O}(Q^2)$ additions to the renormalization 
parameter.  Thus, in this ``pQCD version'' of the  scenario with the flat DA, 
we deal simply with $\varphi_\pi (x)$. 
It does not evolve in the photon-pion transition amplitude, 
so there is  no need to specify  at which scale it is defined.

Furthermore, writing the   square-bracketed term  in Eq.~(\ref{Jr}) as 
$[A(r)-B(r) \ln (Q^2/\mu^2)]$, we can fine-tune the 
coefficient $a$ by taking $a=\exp[-A(r)/B(r)]$ so
as to completely eliminate the one-loop correction. 
Still, the resulting $\mu^2=a Q^2$ will be very small, and 
there will be no evolution   change in
the shape of flat DA.  In other words, in the  pQCD version of  flat DA scenario,
there is no need to consider radiative corrections for the photon-pion 
transition form factor:  they all are absorbed by the
pion wave function.

The photon-pion transition form factor was 
 investigated in Ref.~\cite{Ruiz Arriola:2006ii}
using   large-$N_c$ 
radial Regge model   for resonances 
coupled to $q_1$ and   $q_2$ photons.
The results obtained in this way may be interpreted  
as a model with flat pion DA at low normalization 
point.  In particular, $\sim \log  Q^2 $ 
behavior was obtained for $Q^2 F_{\gamma^*\gamma \pi^0} (Q^2)$ 
in the large-$Q^2$ region.  The authors used  the leading logarithm 
prescription $\phi_\pi (x) \to \phi_\pi (x, Q)$  
to make comparisons with experiment,
i.e. the pion DA  in their approach is  not exactly flat
for large $Q^2$.

When  the pQCD version  of the scenario with flat pion DA 
is  applied to  pion electromagnetic form factor, 
the analysis  of radiative corrections 
is very similar. The conclusion is that 
there is no need to consider the one-gluon-exchange diagram:
the gluon line should be absorbed into the soft wave function.
After that, only the soft contribution remains, 
and the form factor should be calculated 
nonperturbatively. 

\begin{figure}[ht]
\includegraphics[height=1.5in]{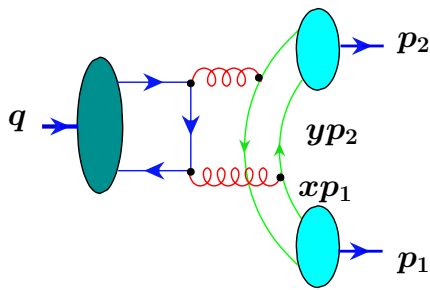}
\caption{Diagram for charmonium decay into two pions:
the gluon lines cannot be absorbed into soft pion 
wave function
}
\label{chipp}
\end{figure}

This does not mean that the flat DA scenario 
excludes the diagrams with  gluon  exchanges for all processes.
Consider charmonium decays into two pions, $\chi_c \to \pi \pi$.
 The two gluons present in the lowest-order diagram cannot be absorbed 
into the pions' wave functions, so this diagram remains.
In pQCD, it produces the same integrals of $\phi (\xi)/(1-\xi^2)$ type
that diverge for flat DA. Thus, one should write 
a more detailed expression involving the $k_\perp$-dependent 
light-front wave functions for both pions,
which is a challenging problem for future studies. 
The description of charmonium decays is  a well-known
success of  CZ approach: if one uses pion DAs close 
to the asymptotic one, 
the theoretical results are well below the 
experimental data.
In case of unmodified propagators,
the   flat scenario gives divergent results for these 
amplitudes, while propagator modification 
brings them down to finite values.
It is  interesting to check if the resulting values are close to CZ ones.

One may argue that 
our  logarithmic or Gaussian models for the lowest-order term
are more or less equivalent to a simple cut-off of the $x$-integral
at $x =M^2/Q^2$  value,  which is essentially larger 
than $x \sim \exp[-1/r]$ values that are responsible for the 
dominant $1/r, 1/r^2, 1/r^3$  terms  in the analysis above.
If one simply  imposes the  cut-off   at $x =M^2/Q^2$ 
in the pQCD expression 
(\ref{14}),  one would get powers of $\ln (Q^2/M^2)$
instead of  of powers of $1/r$, and since $\ln (Q^2/M^2)\lesssim 4$
in our case, the  asymptotically nonleading terms (especially (-9/2)
contribution,
see Eq.(\ref{14})) are essential. 
But it is not clear if a simple $x$ cut-off in  the pQCD expression 
is a correct prescription in the one-loop case.
 In particular, one may  notice that the leading-order
formula (\ref{75A}) of the light-front formalism  can   be 
formally written in terms of pion DA  
$\varphi_\pi (x,\mu)$ taken at $\mu =xQ$ (cf. Eq.~(\ref{phiLB})):
\begin{equation}
F^{\bar qq}_{\gamma^*\gamma \pi^0} 
(Q^2) = \frac{ \sqrt{2}}{3} 
\int_0^1 \frac{dx}{xQ^2} \, \varphi_\pi (x, \mu=xQ) \, 
 .
  \label{75B} \end{equation}
So, if the most important values are $x \sim M^2/Q^2$,
then one should take $\mu \sim M^2/Q$
(which is $\mu \sim Q\,e^{-\ln(Q^2/M^2)}$, compare with 
(\ref{1/r})), i.e. again a very small
value for large $Q$. 
However, to check  if this reasoning extends to the one-loop case, 
 one needs to calculate one-loop corrections
in the light-front formalism keeping the $k_\perp$-dependence
in the perturbative part  
and then convoluting the result with $k_\perp$-dependent
nonperturbative
wave function(s), which is a task going well  beyond
 the scope of the present paper.

\section{Summary}

In this paper we discussed a scenario in which 
pion distribution amplitude is treated as a constant 
for all values of the light-cone momentum fraction $x$.
We indicated that several approaches,
in particular  QCD sum rules and lattice gauge calculations
give the values for the second moment 
$\langle \xi^2 \rangle$ of the 
pion distribution amplitude that are compatible 
with this proposal.
We emphasized that the standard practice 
of building the model pion DAs as a sum of 
two or three  lowest terms  of the Gegenbauer expansion
is  just an assumption.  
Such an assumption, however, excludes flat DAs from the start.
We calculated the photon-pion transition form factor 
using the light-front formula of Lepage and Brodsky 
and incorporating a $\bar q q$ wave function 
that gives flat pion DA and has a rapid 
(exponential, for definiteness) fall-off 
with respect to light-front energy combination
$k_\perp^2/x(1-x)$.
We  demonstrated that the  use of such a wave function
is  numerically equivalent to $1/xQ^2 \to 1/(xQ^2 +M^2)$
modification of the quark propagator, with the parameter
$M^2$  being more than three times larger compared to 
the average square of the valence quark transverse momentum.
The characteristic feature of our result is logarithmic $\sim \ln \, (Q^2/M^2+1)$ growth
with $Q^2$ of the combination
$Q^2 F_{\gamma*\gamma \pi^0} (Q^2) $.
Such a growth is   indicated by recent 
data of {\sc BaBar}  collaboration \cite{Aubert:2009mc}.
In this  respect, it looks  very important 
to   check these results at other facilities, such as BELLE.

\section{Acknowledgments}

I thank V.~M.~Braun and N.~A.~Kivel for bringing 
the results of {\sc BaBar}  experiment to my attention
and stimulating communication.
My special  thanks for discussions are 
due to M.~V.~Polyakov who informed 
me that he independently  reached the conclusion that
recent  {\sc BaBar}  data may be explained 
by a pion DA that does not vanish at the end-points
(see Ref.~\cite{Polyakov:2009je}).

This paper is authored by Jefferson Science Associates,
LLC under U.S. DOE Contract No. DE-AC05-06OR23177. 
The U.S.
Government retains a non-exclusive, paid-up,
irrevocable, world-wide license to publish or reproduce this
manuscript for U.S. Government purposes.

\end{document}